\documentclass[12pt]{article}
\usepackage{amsmath, amssymb}
\usepackage{amsbsy}

\usepackage{hyperref}

\def\half{\textstyle\frac12}
\def\bJ{{\bf J}}

\begin{document}

\begin{center} {\Large \bf
Boundary unitarity and the black hole information paradox\footnote{Essay written for the Gravity Research Foundation 2013 Awards for Essays on Gravitation. Submission date: March 31, 2013.
This essay essentially coincides, apart from a few stylistic changes, with version 1 of the article posted originally at arxiv.org under the title ``Boundary unitarity without firewalls". It is  posted 
as version 2 of that article.
}}
\end{center}

\vskip 5mm
\begin{center} \large
{{Ted Jacobson
}}
\end{center}

\vskip  0.5 cm
{\centerline{\it Maryland Center for Fundamental Physics}}
{\centerline{\it Department of Physics, University of Maryland}}
{\centerline{\it College Park, MD 20742-4111, USA}
{\centerline{\it jacobson@umd.edu}
}

\vskip 1cm

\begin{abstract}{\small
Both AdS/CFT duality and more general reasoning from quantum gravity
point to a rich collection of boundary observables that always evolve unitarily.
The physical quantum gravity states described by these observables
must be solutions of the spatial diffeomorphism and Wheeler-deWitt constraints,
which implies that the state space does not factorize into a tensor product 
of localized degrees of freedom. The ``firewall" argument that unitarity of 
black hole S-matrix implies the presence of a highly excited quantum state
near the horizon is based on such a factorization, hence is not applicable 
in quantum gravity. In fact, there appears to be no conflict between boundary unitarity
and regularity of the event horizon.} 
\end{abstract}


According to low energy effective
field theory on a fixed spacetime background, the 
ground state of a quantum field near a black hole is 
a superposition of maximally entangled pairs, with one member of the pair localized
outside the horizon and the other one inside. Any significant deviation from this structure
would imply that the state is actually not the local ground state but is excited, 
and the energy density in that that excited state would grow exponentially as it is 
followed backwards in time toward the horizon.  This situation is similar 
to the one in primordial cosmology,
which implies that the quantum state during a period of inflationary exponential expansion
of the scale factor must essentially be the de Sitter vacuum, since otherwise at earlier times
the energy density of excitations would have overwhelmed the uniform vacuum energy driving
the inflation in the first place. 

Since regularity of the horizon requires entanglement across the horizon, 
it seems to be in conflict with the hypothesis that black hole evaporation 
is unitary.  It was recently emphasized 
in Ref.~\cite{Almheiri:2012rt} (AMPS) that the reasoning that leads to this conflict
is independent of quantum gravity considerations.
The AMPS argument assumes both that (1) there is a unitary S-matrix 
describing evolution from a pure state of infalling matter to outgoing Hawking radiation, and 
(2) low energy effective quantum field theory (EFT) in the black hole background 
holds to a good approximation beyond some microscopic distance from the horizon.
Then if the initially mixed, thermal Hawking radiation is to be purified by virtue of 
entanglement with later Hawking radiation, it must be that around 
halfway through the evaporation, \emph{at the latest}, emerging Hawking quanta are entangled,
at least somewhat, with earlier Hawking radiation. Such entanglement would imply that those
later quanta are not maximally entangled with their partners inside the horizon, however, which
would mean that the field is not in the local ground state. Tracing this state backward
in time, but remaining within the domain of the low energy EFT, they
infer that there must be a ``firewall" just outside the horizon, that is, a highly excited state.

The firewall conclusion seems highly implausible, basically
since a weak quantum gravity effect should not have a
huge back reaction on a macroscopic object. To avoid this
conclusion, assuming the validity of the AMPS argument, 
one must find a reason to reject at least one of the assumptions. 
For the same reason as the firewall is implausible, 
the semi-classicality assumption (2) seems highly plausible,
and the unitarity assumption (1) seems at first highly implausible,
since information about what falls into a black hole can apparently 
be lost forever to the outside. 

However, many arguments have been advanced arguing 
that black hole evaporation is in fact 
unitary as viewed just from the outside. In my view almost
none of those are convincing. For example, it 
is often stated that 
if information could be lost inside black holes then 
the principles of quantum mechanics would be violated, leading to 
a breakdown in the theory, or that 
the black hole pair creation rate would be far too large, or that 
rampant violations of energy and momentum conservation would occur
in the vacuum. Putting aside all such unconvincing arguments however, there 
remains a powerful argument, coming from AdS/CFT duality,
 in favor of black hole unitarity. 
The argument is simply that the CFT is a UV complete quantum field theory
on a fixed background manifold, hence is ``manifestly" unitary. At the same
time, the CFT is presumably dual to at least enough of the gravity/string theory in
the AdS bulk to capture all asymptotic observables in the theory, such as those
associated with the black hole S-matrix.

Even if the AdS/CFT argument held only in that specific setting, 
it would be enough to pose the AMPS puzzle.
But in fact Marolf \cite{Marolf:2008mf} has argued that the essential reason for the AdS/CFT result
carries over more generally to any diffeomorphism invariant, UV complete quantum theory
with an asymptotic region in which an algebra of observables can be defined. 
His point is that in such a theory, the Hamiltonian is a surface integral in the 
asymptotic region, which I will call ``the boundary". More precisely, the Hamiltonian also
contains a volume integral of combinations of the diffeomorphism constraints,
but those act trivially\footnote{Actually it might just be that matrix elements of the constraints
between physical states vanish, but that is probably good enough to make the argument.} 
on any physical state in the Hilbert space (according to Dirac
quantization of a constrained system). Hence the algebra of boundary observables
evolves unitarily in time into itself, and this means that no boundary information can ever be lost.
In the asymptotically flat case, the boundary algebra would have 
to be located at null infinity or perhaps at spacelike infinity \cite{spi} (rather than at the point at
spacelike infinity $i^0$) if a nontrivial notion 
of time evolution were to be present. Although the details have not been worked out,
this generalization of the idea of boundary unitarity is at least plausible.
For the present discussion I will assume its validity, but if necessary the
arguments could be restricted to the AdS setting.

Now let's consider the implications of this ``boundary unitarity".  
Note first that it implies that {\it boundary information is never lost, not even temporarily}. 
This is quite unlike the situation considered by AMPS. They contemplate a process
in which a certain amount of early time Hawking radiation arrives, far from the black hole, in a 
mixed quantum state, and therefore needs to be purified later by the late time Hawking radiation,
in order that, in the end, the final out-state will be unitarily related to the in-state.
Boundary unitarity would never allow such a scenario if the Hawking quanta are described by
boundary observables, and it has nothing to say about them if they are \emph{not}. 
Hence I will assume that they are.

Boundary unitarity then means that both a Hawking quantum 
and the degrees of freedom with which it is 
entangled -- whatever they are -- are described
within the boundary algebra. This much seems obvious and inevitable in AdS/CFT and, more generally
according to Marolf's argument, in any UV complete diffeomorphism invariant theory. 
For this reason I think that the AMPS assumption (1) is unmotivated, since there is 
no reason to presume that, taken by itself, the Hawking radiation is in a pure state.  
I propose that it should be replaced by 
the different assumption (1b) of boundary unitarity. That is, the question that should be 
considered is not whether the Hawking radiation can manage to be self-purifying in the end without 
engendering a firewall, but whether it can be \emph{continuously pure} when taken together with 
the rest of the asymptotically observable physical state space.

It might appear that the AMPS argument carries over to 
assumption (1b), because continuous purity seems to require that
a Hawking quantum be entangled with exterior degrees of freedom, so that it could not  
be maximally entangled with its partner behind the horizon. According to the AMPS 
reasoning this would require 
a firewall right from the beginning of the Hawking emission. But there is a deeper aspect of the
change from assumption (1) to (1b). Namely, (1b) refers only to gauge-invariant states
in the physical quantum gravity Hilbert space, which are annihilated by the diffeomorphism 
constraints (spatial diffeomorphism invariance and the Wheeler-de Witt equation), 
whereas (1) refers to an approximate concept of quantum field theory states in 
a background geometry, and assumes the Hilbert space is a tensor product of
exterior and interior state spaces. But this factorization is certainly not a feature of the
physical Hilbert space of quantum gravity states.
The representative of the Hawking quantum's partner at the boundary 
is {\it not} another factor of the Hilbert space. In fact, the ``partner behind the horizon" is a fictional character in
full quantum gravity, which on account of diffeomorphism invariance has no such localizable degrees of freedom. 

So how then is the partner, i.e. the purifier of a Hawking quantum, encoded at the boundary? 
I suppose it must be in the gravitational imprinting of the quantum 
fluctuation that corresponds to each particular realization of the Hawking process. This imprinting
is conveyed via the diffeomorphism constraints (in particular the Wheeler-de Witt equation), 
and reaches to infinity. 
If interior information were
{\it copied} onto another factor in the Hilbert space, this cloning of quantum information 
would violate the linearity of quantum mechanics~\cite{Wootters:1982zz}.
But this gravitational imprinting is not a duplication of information because, modulo the constraints, 
the Hilbert space is not a tensor product of interior and exterior degrees of freedom, and 
the independent observables are fewer in number than semiclassical
reasoning leads us to believe \cite{Giddings:2005id,Papadodimas:2012aq}.

To illustrate how the quantum mechanics can work in this way it is helpful to consider 
a simple model consisting of spins. Suppose the system consists of four spin-$\half$ 
degrees of freedom, for which the usual Hilbert space $\half\otimes\half\otimes\half\otimes\half$
would be $2^4=16$ dimensional.  To model the diffeomorphism 
constraint, let us add the requirement that the ``physical" states consist only of the singlets, i.e.\ 
the states with vanishing total angular momentum. To find the singlets 
we can decompose the unconstrained tensor product Hilbert space into irreducible spin representations:
\[(\half\otimes\half)\otimes(\half\otimes\half) = (0\oplus1)\otimes(0\oplus1)= 
0\oplus1\oplus1\oplus(1\otimes1).\] 
The last term is equal to $0\oplus 1\oplus 2$, so there are two singlets. 
The singlet subspace is not a tensor product. Moreover, 
although these singlet states are collective in the original unconstrained
degrees of freedom, they can be distinguished by observables that act only on
a subspace of the unconstrained Hilbert space. 
For example, let $\bJ_1$ and $\bJ_2$ denote the angular momentum
operators of the second pair. Then the eigenvalues of $(\bJ_1+\bJ_2)^2$
are 0 and 1, which label the first and second singlets respectively. 
The same kind of measurement could have been made instead on the first
pair, or even on a different pairing, and still it would serve to fully
measure the state within the physical Hilbert space.
In this spin model, there is just one constraint,
while in the quantum gravity case there are four constraints per spatial point. 
This serves as a reminder of how different the physical state space is from 
the unconstrained Hilbert space.
 
The fact that the physical Hilbert space does not factorize into degrees of freedom 
inside and outside a horizon, and that the states have a global nature, does not mean it is impossible 
to make measurements of effectively local degrees of freedom. One could lower a probe to 
a back hole horizon to interact with the ``local" degrees of freedom. 
However, in this case, the full quantum state,
including the ``local" measuring device, must satisfy the diffeomorphism constraints, and
the presence of the measuring device changes the nature of the physical
states. 

My story is obviously lacking in
the essential details of exactly what is the nature of the physical states, and 
how the ``partners" and more generally any other
degrees of freedom we would semiclassically think of as residing behind the 
horizon are encoded in the boundary.
In the AdS/CFT setting however, Ref.~\cite{Papadodimas:2012aq} proposes a concrete answer to 
this question, at least for the partners. That paper is in a sense dual to this one.
Rather than invoking the constraints of bulk quantum gravity, the authors start from the CFT
side and show how, associated with an approximately thermal but \textit{pure} state, one can 
use the Schmidt decomposition of the state to construct 
an algebra of operators corresponding to the partners behind a horizon in the hologram.
Based on this construction, they are able to be rather precise about the sense in which 
the interior degrees of freedom are redundant with exterior degrees of freedom. 

In conclusion, when the logical structure of the quantum gravity theory is fully taken into account, 
the firewall paradox is resolved. In particular, 
the distinction between the field theory Hilbert space on a background spacetime
and the physical state space is crucial for resolving the paradox. 
This should be expected, since only for physical states does the
Hamiltonian reduce to a boundary term, and therefore only for such states do we
have any reason to expect boundary unitarity to hold.

\section*{Acknowledgments}
This research was supported in part by NSF grant PHY-0903572,
and by the Kavli Institute for Theoretical Physics through NSF grant 
PHY11-25915.

\end{document}